\documentclass[aps,pra,preprint,superscriptaddress,longbibliography]{revtex4-1}

\usepackage{amsmath}
\usepackage{amsthm}
\usepackage{amsfonts}
\usepackage{amssymb}
\usepackage{xcolor,graphicx}
\usepackage{tikz}
\usepackage{color}
\usepackage[pdftex]{hyperref} %para ligar las referencias
\usepackage{longtable}
\usepackage{array}
\usepackage{dsfont}

\begin{document}
	
%--------------------------------------------------------------------------------
	\title{Isospectral and square root Cholesky photonic lattices}
	
	\author{P.~I. Martinez Berumen}
	\email[e-mail: ]{A00818058@itesm.mx}
	\affiliation{Tecnologico de Monterrey, Escuela de Ingenier\'ia y Ciencias, Ave. Eugenio Garza Sada 2501, Monterrey, N.L., Mexico, 64849}

	\author{B. M. Rodr\'iguez-Lara}
	\email[e-mail: ]{bmlara@tec.mx}
	\affiliation{Tecnologico de Monterrey, Escuela de Ingenier\'ia y Ciencias, Ave. Eugenio Garza Sada 2501, Monterrey, N.L., Mexico, 64849}
	
	\date{\today}
	
	\begin{abstract}
	Cholesky factorization provides photonic lattices that are the isospectral partners or the square root of other arrays of coupled waveguides.
	The procedure is similar to that used in supersymmetric quantum mechanics. 
	However, Cholesky decomposition requires initial positive definite mode coupling matrices and the resulting supersymmetry is always broken. 
	That is, the isospectral partner has the same range than the initial mode coupling matrix.
	It is possible to force a decomposition where the range of the partner is reduced but the characteristic supersymmetric intertwining is lost.
	As an example, we construct the Cholesky isospectral partner and the square root of a waveguide necklace with cyclic symmetry.
	We use experimental parameters from telecommunication C-band to construct a finite element model of these Cholesky photonics lattices to good agreement with our analytic prediction.
	\end{abstract}
	
	%\pacs{05.45.Mt, 42.50.Ct, 42.50.Mn, 73.43.Nq}
	
	\maketitle
%--------------------------------------------------------------------------------
%%%%%%%%%%%%%%%%%%%%%%%%%%  body  %%%%%%%%%%%%%%%%%%%%%%%%%%
\section{Introduction}

The optical analogy of supersymmetric quantum mechanics (SUSY QM)  can be traced back to planar waveguides with elliptic transversal index profile, where the paraxial approximation provides exact SUSY that breaks for non-paraxial fields \cite{Chumakov1994p51}.
Theoretical curiosity gave place to practical applications with the proposal to use SUSY as a tool for the synthesis of optical structures with particular spectral properties in both bulk and discrete optics \cite{Miri2013p233902}.
In particular, discrete SUSY photonic lattices may serve in optical communications providing multiplexing schemes \cite{ Miri2014p6130}, mode selection \cite{Midya2018p3758}, optical intersections \cite{Longhi2015p463}, Bragg grating filters \cite{Longhi2015p045803}, and mode conversion \cite{Heinrich2014p3698,Miri2014p89}, to mention a few examples.

The analogy between the wave equation in the paraxial approximation and the Schr\"odinger equation allows using standard SUSY QM techniques \cite{Cooper1995p267}.
For example, the Darboux transformation of an optical analogue to the Hamiltonian $\hat{H}_{1} =-(\hbar^2/2m) (d^2 / dx^2 ) + V_{1} (x) \equiv \hat{A} \hat{A}^{\dagger}$ to produce an isospectral partner $\hat{H}_{2} =-(\hbar^2/2m) (d^2 / dx^2 ) + V_{2}(x) \equiv \hat{A}^{\dagger} \hat{A}$.
The effective potentials, proportional to the square of refractive index distributions, are related by a super potential $W(x)$ that solves Riccati equations $V_{1}(x)=W^2(x)+(\hbar/\sqrt{2m})W'(x)$, $V_{2}(x)=W^2(x)-(\hbar/\sqrt{2m})W'(x)$ and allows writing the operators $\hat{A} = -(\hbar / 2m) (d^2 / dx^2 ) + W(x)$ and $\hat{A}^\dagger =(\hbar / 2m) (d^2 / dx^2 ) + W(x)$.
This technique is commonly used to design optical systems \cite{Longhi2015p045803,Longhi2010p032111}.
The analysis is done for an infinite dimension device that is cut off to a size large enough to see the desired effects in real world applications \cite{Longhi2010p032111,Longhi2015p045803,Longhi2015p463,Miri2013p233902, Miri2014p6130, Miri2014p89,Midya2018p3758,Midya2018p4927,Heinrich2014p3698,Teimourpour2016p33253,Smirnova2019p1120,Hokmabadi2019p623,Midya2019p363,Zuniga2014p987}.
On the other hand, it is possible to work with finite dimensional optical devices and show SUSY with different Witten indices by addition of $\mathcal{PT}$-symmetry \cite{ElGanainy2012p033813}.
This has inspired the use of optical lattices and their superpartners to desing laser arrays by the addition of gain and loss following different seeding patterns \cite{ElGanainy2015p033818,Teimourpour2016p33253,Midya2019p363,Smirnova2019p1120,Hokmabadi2019p623}.
Factorization methods from linear algebra are a practical tool in some of these designs \cite{ElGanainy2015p033818,Zhong2019p1240,Smirnova2019p1120,Heinrich2014p3698,Miri2013p233902,Miri2014p6130,Hokmabadi2019p623,Midya2019p363}.

Our research program advocates the use of abstract symmetries to optimise optical design processes \cite{Rodriguez2018p244}.
For example, it is possible to construct SUSY photonic lattices partners that have semi-infinite dimension using the special unitary algebra $su(1,1)$ as underlying symmetry \cite{Zuniga2014p987,RodriguezLara2014p1719}. 
While the closed form analysis is done in infinite dimensions, large arrays of the order of hundred of elements follow the analytic predictions.
It is also possible to construct SUSY partners for finite dimensional lattices using, for example, an underlying $su(2)$ symmetry \cite{Teimourpour2016p372}.
In discrete optical systems described by coupled mode theory, Cholesky factorization is a helpful linear algebra tool to decompose the mode-coupling matrix \cite{Zhong2019p1240,Heinrich2014p3698,Miri2013p233902} and, then, use the particular modes as seed to design, for example, parity anomaly lasers \cite{Smirnova2019p1120}.

In the following, we review Cholesky factorization of positive definite real symmetric matrices and its relation with the properties expected from standard SUSY QM with Witten index two \cite{Ramond1971p2415,Neveu1971p86,Witten1981p513,Lahiri1990p1383,Cooper1983p262}.
We show that this approach provides us with isospectral and square root broken SUSY partners, Sec. \ref{sec:Sec2}.
Then, we use waveguide necklaces with an underlying cyclic group $Z_{N}$ symmetry as the original partner to construct practical examples of broken SUSY partners.
In particular, we provide an analytic isospectral partner for a two-waveguide necklace and a square root partner for a four-waveguide necklace.
We compare our theoretic predictions with numeric finite element modelling simulation based on experimental parameters from laser inscribed realizations, Sec. \ref{sec:Sec3}.
In Section \ref{sec:Sec4}, we discuss the fact that it is possible to force a pseudo zero-energy mode. 
The result is a viable optical system that shows the spectral characteristics but is not exact SUSY as the intertwining relation breaks.
We close with a summary and our conclusion, Sec. \ref{sec:Sec5}.

%%%%%%%%%%%%%%%%%%%%%%%%%%%%%%%%%%%%%%%%%%%%%%%%%%%%%%%%%%%%%%%%%%%%%%%%%%%
\section{Cholesky lattices} \label{sec:Sec2}
%%%%%%%%%%%%%%%%%%%%%%%%%%%%%%%%%%%%%%%%%%%%%%%%%%%%%%%%%%%%%%%%%%%%%%%%%%%

Coupled mode theory simplifies the description of electromagnetic field modes propagating though arrays of coupled waveguides \cite{McIntyre1973p63}.
Instead of describing polarized localized spatial field modes at each waveguide, $E_{j} = \mathcal{E}_{j} \Psi(\mathbf{r}) \hat{\epsilon}$, it provides an approximation,
\begin{eqnarray}
i \partial_{z} \mathbf{E} = \mathbf{M} \mathbf{E},
\end{eqnarray}
for the dynamics of the complex field amplitudes summarized in the amplitude vector with $j$-th component $\mathbf{E}_{j} = \mathcal{E}_{j}$.
The diagonal terms of the mode-coupling matrix provide information about the propagation constant of localized field modes, $M_{ii} = \beta_{i} > 0$, and the off-diagonal ones of the coupling strength between modes localized in pairs of waveguides, $M_{ij} =  M_{ji} = g_{ij} >0$.
Usually, nearest neighbours are the strongest coupled and a standard approximation is to neglect high order neighbors.  
In the optical and telecommunication regimes, the propagation constants are at least three orders of magnitude larger than the coupling strengths.
Under these circumstances, the mode coupling matrix is positive-definite.

Cholesky factorization decomposes a positive-definite Hermitian matrix,
\begin{eqnarray}
 	\mathbf{M} = \mathbf{A} \, \mathbf{A}^{\dagger},
\end{eqnarray}
into the product of positive definite lower triangular matrix $\mathbf{A}$ and its conjugate transpose $\mathbf{A}^{\dagger}$; herein, we call these Cholesky matrices.
This suggests the use of SUSY QM ideas to construct the isospectral partner of our mode coupling matrix. 
Let us define a new pair of extended Cholesky matrices, 
\begin{eqnarray}
\mathbf{Q} = \left( \begin{array}{cc} 0 & 1 \\ 0 & 0 \end{array} \right) \otimes \mathbf{A} ~\mathrm{ and }~ \mathbf{Q}^{\dagger} = \left( \begin{array}{cc} 0 & 0 \\ 1 & 0 \end{array} \right) \otimes \mathbf{A}^{\dagger},
	\label{eq:superchargeOperators}
\end{eqnarray}
that are nilpotent by construction, $\mathbf{Q}^{2} = \mathbf{Q}^{\dagger 2} = 0$.
In consequence, these two matrices commute,
\begin{eqnarray}
\left[ \mathbf{H}, \mathbf{Q} \right] =  \left[ \mathbf{H}, \mathbf{Q}^{\dagger} \right] = 0,
\end{eqnarray}
with a new block diagonal matrix,
\begin{eqnarray}
\mathbf{H} = \mathbf{Q} \, \mathbf{Q}^{\dagger} + \mathbf{Q}^{\dagger} \, \mathbf{Q} = \left( \begin{array}{cc} \mathbf{M} & 0 \\ 0 & \mathbf{P} \end{array} \right),
\end{eqnarray}
that has our mode coupling matrix $\mathbf{M}$ and a new matrix,
\begin{eqnarray}
\mathbf{P} = \mathbf{A}^{\dagger}  \, \mathbf{A},
\end{eqnarray}
that we call its partner, in the main diagonal. 
It is straightforward to show a matrix intertwining relation, 
\begin{eqnarray}
\mathbf{Q}^{\dagger} \, \mathbf{H}_{M}  =  \mathbf{H}_{P} \, \mathbf{Q}^{\dagger},
\end{eqnarray}
where we define expanded mode coupling and partner matrices, $\mathbf{H}_{M} = \mathbf{Q} \, \mathbf{Q}^{\dagger}$ and $\mathbf{H}_{P} = \mathbf{Q}^{\dagger} \, \mathbf{Q}$, in that order.
It is possible to construct the normal modes of the extended partner matrix starting from those of the extended coupling matrix,
\begin{eqnarray}
\mathbf{H}_{M} \, \mathbf{m}_{j} = \mu_{j} \, \mathbf{m}_{j},
\end{eqnarray}
and multiply them by $\mathbf{Q}^{\dagger}$ from the left, 
\begin{eqnarray}
\mathbf{Q}^{\dagger} \, \mathbf{H}_{M} \, \mathbf{m}_{j}&=& \mu_{j}~ \mathbf{Q}^{\dagger} \, \mathbf{m}_{j},
\end{eqnarray}
to use the matrix intertwining relation,
\begin{eqnarray}
\mathbf{H}_{P} ~ \mathbf{Q}^{\dagger} \, \mathbf{m}_{j}  = \mu_{j} ~ \mathbf{Q}^{\dagger} \, \mathbf{m}_{j},
\end{eqnarray}
and obtain the extended partner matrix normal modes,
\begin{eqnarray}
\mathbf{H}_{P} \, \mathbf{p}_{j} = \mu_{j} \mathbf{p}_{j}, \, \mathrm{ with }~  \mathbf{p}_{j} = \mathbf{Q}^{\dagger} \,\mathbf{m}_{j}.
\end{eqnarray}
The extended matrix has identical spectrum as long as $\mathbf{Q}^{\dagger} \mathbf{m}_{j} \neq 0$.
Cholesky factorization provides positive definite extended matrices.
In consequence, this method always provides isospectral partners. 

We keep borrowing from SUSY QM and construct a pair of Hermitian matrices, 
\begin{eqnarray}
\mathbf{H}_{X} = \mathbf{Q}^{\dagger} + \mathbf{Q}, ~~\mathrm{and}~~ \mathbf{H}_{Y} = -i \left( \mathbf{Q}^{\dagger} - \mathbf{Q} \right),
\end{eqnarray}
that are the square root of the previous diagonal matrix, 
\begin{eqnarray}
\mathbf{H}_{X}^{2} = \mathbf{H}_{Y}^{2} = \mathbf{H},
\end{eqnarray} 
and share normal modes with it, 
\begin{eqnarray}
\mathbf{H}_{X} ~ \mathbf{x}_{j} = x_{j} ~ \mathbf{x}_{j}, ~\mathrm{and}~  \mathbf{H} ~ \mathbf{x}_{j} = x_{j}^{2} ~ \mathbf{x}_{j}.
\end{eqnarray}
These modes are doubly degenerate for the diagonal matrix $\mathbf{H}$ as we can define some general mode,
\begin{eqnarray}
\mathbf{v}_{j} = \mathbf{H}_{Y}~\mathbf{x}_{j},
\end{eqnarray}
and realize that it is also an eigenvalue of the new matrix,
\begin{eqnarray}
\mathbf{H}_{X} \mathbf{v}_{j} = -x_{j} \mathbf{v}_{j}
\end{eqnarray} 
where we used the fact that $\left\{\mathbf{H}_{X}, \mathbf{H}_{Y}\right\} = \mathbf{H}_{X} \mathbf{H}_{Y} + \mathbf{H}_{Y} \mathbf{H}_{X} = 0$ leads to the relation $\mathbf{H}_{X} \mathbf{H}_{Y} = -\mathbf{H}_{Y} \mathbf{H}_{X}$.
This eigenvalue equation implies that the spectrum of the block diagonal matrix $\mathbf{H}$ is doubly degenerate and the spectrum of the block anti-diagonal matrix $\mathbf{H}_{X}$ has paired eigenvalues $\pm x_{j}$.

Before moving forward to practical examples, we want to stress that the Cholesky decomposition of Hermitian positive definite mode coupling matrices $\mathbf{M}$ provides isospectral partners $\mathbf{P}$.
Thus, the block diagonal matrix $\mathbf{H}$ has a doubly degenerate spectrum and its square root matrix $\mathbf{H}_{X}$ has paired spectrum.
Thanks to the fact that the mode coupling matrix diagonal terms are larger than the off-diagonal, we can find a sequence of isospectral partners and square root matrices just by decomposing,
\begin{eqnarray}
\mathbf{M} = \alpha \mathbf{1} + \mathbf{M}_{\alpha}.
\end{eqnarray}
As long as the new effective mode coupling matrix $\mathbf{M}_{\alpha}$ is positive definite, we can find isospectral and square root matrices of $\mathbf{M}_{\alpha}$ for each parameter $\alpha$ that might be experimentally viable or not. 

%%%%%%%%%%%%%%%%%%%%%%%%%%%%%%%%%%%%%%%%%%%%%%%%%%%%%%%%%%%%%%%%%%%%%%%%%%%
\section{Waveguide necklaces} \label{sec:Sec3}
%%%%%%%%%%%%%%%%%%%%%%%%%%%%%%%%%%%%%%%%%%%%%%%%%%%%%%%%%%%%%%%%%%%%%%%%%%%

In order to provide a working example, we study a so-called waveguide necklace composed by $N$ identical cores equidistantly distributed on a circle of radius $r$.
The spectrum of these arrays is straightforward to calculate including couplings of all orders~\cite{Jaramillo2019p515}.
We assume a weakly coupled necklace described by the mode coupling matrix,
\begin{eqnarray}
	\left[ \mathbf{M}(\beta_{0},g) \right]_{i,j} = \beta_{0} \delta_{i,j}+g(\delta_{i,j+1}+\delta_{i+1,j}),
\end{eqnarray}
where the propagation constant of the localized modes at each waveguide is $\beta_{0}$, the coupling strength between first neighbors is $g$, and the addition in Kronecker delta subindices is modulus $N$ such that $N + k \equiv \mathrm{mod}_{N}(N+k) = k$.
The spectrum is positive definite,
\begin{equation}
\beta_{j} = \beta_{0} + g \left\{ 
\begin{array}{ll} 
	1 & N=2,\\
 2 \cos \frac{(j-1)}{m}\pi  + (-1)^{j-1},  & N=2m,  \\
 2\cos \frac{2(j-1)}{2m+1}\pi  , & N=2m+1,
\end{array} \right.
\label{eq:spectrumNecklaces}
\end{equation}
and has $m$ duplicated elements with one (two) non-duplicated values for odd (even) dimension.
The spectrum elements with minimum value,
\begin{equation}
\beta_{\mathrm{min}} = \beta_0 - g \left\{ 
\begin{array}{ll} 
  1,  & N=2,  \\
  2, & N=2m,  \\
  2 \cos \frac{2m}{2m+1} \pi , & N=2m+1,
\end{array} \right.
\label{eq:minimumEigenvalueNecklaces}
\end{equation}
suggest the decomposition,
\begin{eqnarray}
	\mathbf{M}(\beta_{0},g) = (\beta_{0} - \epsilon) \mathbf{1} + \mathbf{M}(\epsilon,g), ~~ \epsilon > \beta_{0}-\beta_{\mathrm{min}},
\end{eqnarray}
to construct any given Cholesky isospectral and square root matrices by focusing on just the positive definite reduced coupled mode matrix $\mathbf{M}(\epsilon,g)$.

\begin{figure}[htbp]
	\includegraphics{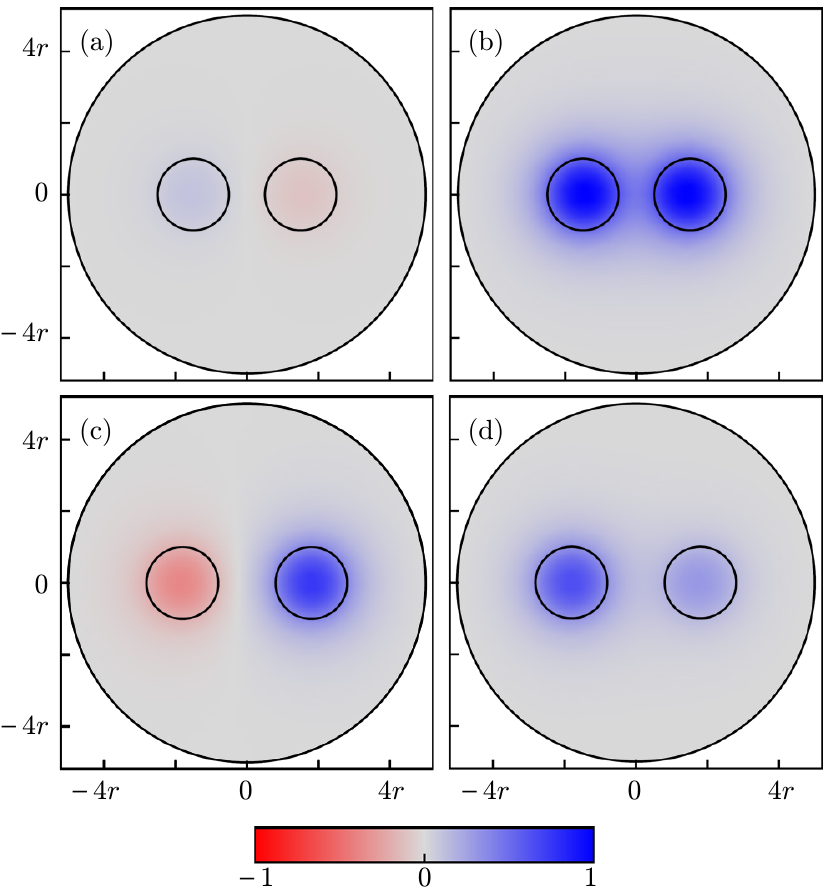}
	\caption{Finite element modelling of normal modes (a)-(b) of a two-waveguide necklace and (c)-(d) its isospectral partner. The propagation constant is (a)-(c) $\beta_{a} = 5.876\,48 \times 10^{6}~ \mathrm{rad/m}$ for assymetric and (b)-(d) $\beta_{s} = 5.877\,31 \times 10^{6}~ \mathrm{rad/m}$ for symmetric modes; see text for more detail.}\label{fig:Fig1}
\end{figure}

%%%%%%%% %%%%%%%% %%%%%%%% %%%%%%%% %%%%%%%% %%%%%%%% 
\subsection{Isospectral partner example} 

Let us start from the simplest analytically solvable example, two waveguides with reduced coupled mode matrix,
\begin{eqnarray}
\mathbf{M}(\epsilon,g) = \left( \begin{array}{cc} \epsilon & g \\ g & \epsilon \end{array} \right), \qquad \epsilon>g
\label{eq:differentWaveguides}
\end{eqnarray}
with eigenvalues,
\begin{eqnarray}
		\lambda_1 = \epsilon-g ~~ \mathrm{and} ~~
		\lambda_2 = \epsilon+g
\end{eqnarray}
and Cholesky decomposition,
\begin{eqnarray}
	\mathbf{A} = \left(
\begin{array}{cc}
 \sqrt{\epsilon } & 0 \\
 \frac{g}{\sqrt{\epsilon }} & \sqrt{\frac{\epsilon^2 - g^2}{\epsilon }} \\
\end{array}
\right),
\end{eqnarray}
yielding a partner mode coupling matrix,
\begin{eqnarray}
	\mathbf{P}(\epsilon,g) = \left(
\begin{array}{cc}
  \frac{\epsilon^{2} + g^2}{\epsilon } & g\sqrt{\frac{ \epsilon^2-g^2}{\epsilon^2}} \\
 g\sqrt{\frac{\epsilon^2-g^2}{\epsilon^2}} & \frac{\epsilon^{2} - g^2}{\epsilon } \\
\end{array}
\right),
\label{eq:superPartnerHamiltonian}
\end{eqnarray}
isospectral to the original matrix $\mathbf{M}(\epsilon,g)$ with a different experimental arrangement.
The diagonal elements point to a $2g^2/\epsilon$ difference between the propagation constants of the localized modes in the waveguides and their coupling constant is smaller than the original partner.
We introduce the propagation constant difference into our design by controlling the transverse area or the refractive index of the waveguide cores and the smaller coupling constant by separating the waveguides. 

As a practical example, we use two cylindrical waveguides of radius $r =4.5~\mu\mathrm{m}$, with core and cladding refractive indices $n_{\mathrm{co}}=1.4479$ and $n_{\mathrm{cl}}=1.444$, respectively, and separation between core centres of $3\,r$ . 
In the telecomm C-band, $\lambda = 1550 ~\mathrm{nm}$, these leads to localized mode propagation constants and coupling strength $\beta = 5.876\,42 \times 10^6 \,\mathrm{rad/m}$ and $g = 416.193 \,\mathrm{rad/m}$, in that order. 
We propose a value of $\epsilon = 1.1 g$ to construct a partner mode coupling matrix.
This implies a difference between the effective localized propagation constants of $\Delta \beta = 756.715 \,\mathrm{rad/m}$ and coupling strength $g = 173.385 \,\mathrm{rad/m}$.
The difference in propagation constants corresponds to an increment of $7.496 \times 10^{-3} \,\%$ in the refractive index of one of the waveguides that is reasonable with changes in the writing speed for laser inscribed setups \cite{Davis1996p1729,Shah2005p1999, Blomer2006p2151,Szameit2006,Heinrich2009,Szameit2010p163001}. 
The new coupling strength implies a separation of $3.606\,060 \,r$ between waveguide cores, Fig. \ref{fig:Fig1}. 
The analytic effective propagation constants for the asymmetric and symmetric normal modes are $\beta_{a}= 5.876 \,003 \, \times 10^{6} \mathrm{rad/m}$ and $\beta_{s}=5.876 \, 835 \, \times 10^{6} \,\mathrm{rad/m}$ and the finite element model simulation provides $\beta_{a}=5.859\,239 \, \times 10^{6} \,\mathrm{rad/m}$ and $\beta_{s}=5.860 \,186\, \times 10^{6} \,\mathrm{rad/m}$ for $\mathbf{M}$, and $\beta_{a}=5.859\, 626 \, \times 10^{6} \,\mathrm{rad/m}$ and $\beta_{s}=5.860 \, 143 \, \times 10^{6} \,\mathrm{rad/m}$ for $\mathbf{P}$, that are within $0.3 \%$ of the predicted values.

\subsection{Square root example}

A waveguide necklace with four elements $N=4$ described by the following real symmetric, positive definite reduced coupled mode matrix, 
\begin{eqnarray}
\mathbf{M}(\epsilon, g) = 
\left( \begin{array}{cccc} 
\epsilon 	& g 		& 0 		& g \\
g 			&\epsilon 	& g 		& 0  \\
0			& g 		& \epsilon	& g  \\
g			& 0			& g			& \epsilon
\end{array}\right).
\end{eqnarray}
with the restriction  $\epsilon > 2g$ has real positive spectrum $\{\epsilon -2g, \epsilon, \epsilon, \epsilon + 2g \}$ with corresponding orthonormal modes $\mathbf{m}_{1} = (-1,1,-1,1)/2$, $\mathbf{m}_{2} = (0,-1,0,1)/\sqrt{2}$, $\mathbf{m}_{3} = (-1,0,1,0)/\sqrt{2}$ and $\mathbf{m}_{4} = (1,1,1,1)/2$ independent of the system parameters $\{ \epsilon, g \}$.
Its associated Cholesky matrix,
\begin{eqnarray}
\mathbf{A} = \left( \begin{array}{cccc} 
\sqrt{\epsilon} 	& 0 		& 0 		& 0 \\
\frac{g}{\sqrt{\epsilon}}	& \sqrt{\frac{\epsilon^2 - g^2}{\epsilon}} 	& 0 		& 0  \\
0					& g\sqrt{\frac{\epsilon}{\epsilon^2 - g^2}} 		& \sqrt{\frac{\epsilon(\epsilon^2 - 2 g^2)}{\epsilon^2 - g^2}}	& 0  \\
\frac{g}{\sqrt{\epsilon}}	& \frac{-g^2}{\sqrt{\epsilon (\epsilon^2 - g^2)}}			& \frac{g \epsilon^2}{\sqrt{2 g^4 \epsilon - 3 g^2 \epsilon^3 + \epsilon^5}}			& \sqrt{ \frac{\epsilon (\epsilon^2 - 4 g^2)}{\epsilon^2 - 2g^2}}
\end{array}\right), \nonumber \\ 
\end{eqnarray}
has one negative element. 
This is not an issue as it is possible to falsify negative couplings using additional elements \cite{Keil2016p312903}.

\begin{figure}[htbp]
	\centering
	\includegraphics{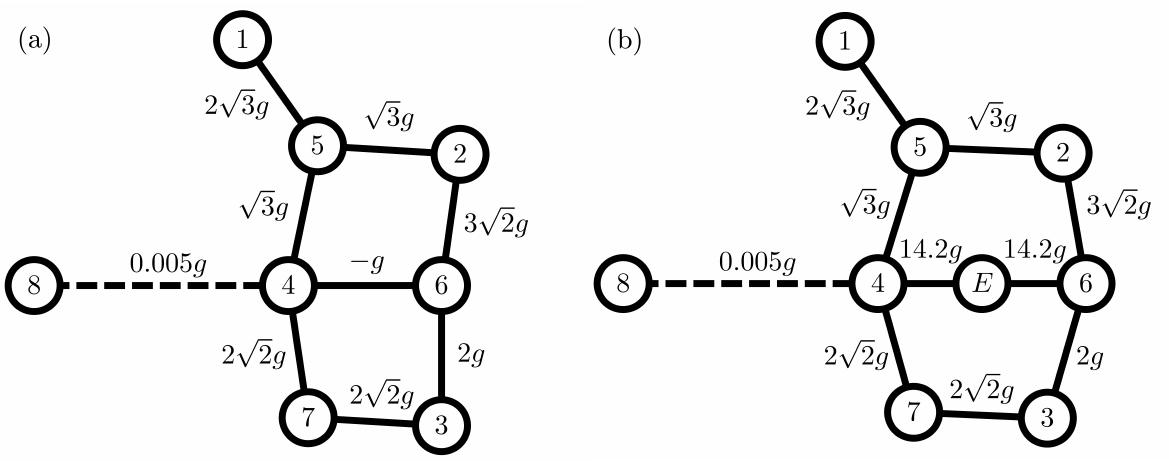}
	\caption{Sketch of (a) the square root Cholesky lattice associated to a four element necklace, note the negative coupling, and (b) its nine waveguide realization; see text for more detail.}
	\label{fig:Fig2}
\end{figure}

Our square root lattice requires an array of eight coupled waveguides with one negative coupling, Fig. \ref{fig:Fig2}(a).
We falsify it using nine waveguides, Fig. \ref{fig:Fig2}(b), that share an effective core radius $r= 4.5 ~\mu\mathrm{m}$ and cladding material with refractive index $n_{\mathrm{cl}}= 1.444$ as before.
The refractive index of sites four and six is $n_{4} = n_{6} = 1.447\,901$, the rest share the index $n_{i} = 1.447\,900$ with $i=1,2,3,5,7,8$, the auxiliary waveguide has an index $n_{E} = 1.448\,094$. 
The distances $d_{ij}$ between the $i$-th and $j$-th waveguides are
$\{d_{15}, d_{25}, d_{26}, d_{36}, d_{37}, d_{45},  d_{47}, d_{4E}, d_{6E}, d_{18} \}= \{5.00224, 5.5, 5.10528, 5.39646, 5.14751, 5.5,\\  5.14751, 4, 4, 9.79616 \} \,r$
with corresponding coupling strengths 
$\{g_{15}, g_{25}, g_{26}, g_{36}, g_{37}, g_{45}, g_{47},\\ g_{4E}, g_{6E}, g_{18} \}= \{24.1136, 12.0568, 20.883, 13.922, 19.6887, 12.0568, 19.6887, 98.8544, 98.8544,\\ 0.0341018 \} \,\mathrm{rad/m}$.

\begin{figure}[htbp]
	\includegraphics{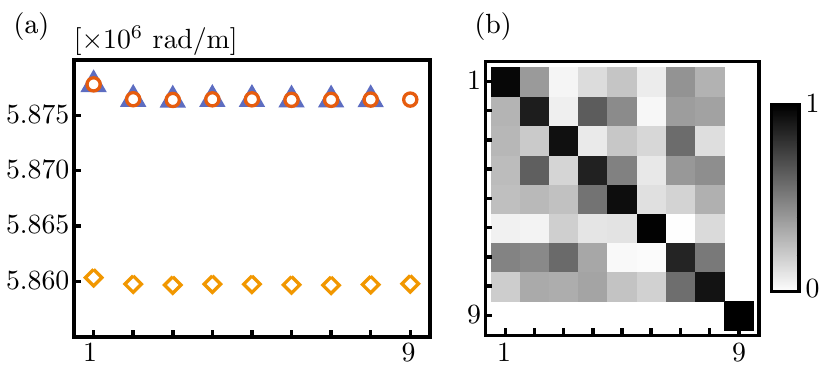}
	\caption{ (a) Propagation constant  from the analytic Cholesky square root array with negative coupling strength (triangles), its analytic nine-waveguide realization (circles) and its finite element model simulation (diamonds). (b) Fidelity overlap between analytic and numerical normal modes.}
	\label{fig:Fig3}
\end{figure}

Figure \ref{fig:Fig3}(a) compares the propagation constants obtained from the eight waveguide array with a negative coupling provided by the analytic Cholesky factorization in triangles, its nine waveguide array realization where all coupling strengths are positive in circles, and the numeric result from finite element modelling in diamonds.
The average relative error between the nine and eight waveguide arrays is of the order of $ \left( 3.747 \pm 8.570 \right) \times 10^{-5} \%$ while that between the numerical finite element model and the analytic eight waveguide array is of the order $ \left( 2.855 \pm 0.044 \right) \times 10^{-1} \%$.
In addition, we use the fidelity overlap,
\begin{eqnarray}
\mathcal{F} = \vert \mathbf{a}_{j}^{\ast} \cdot \mathbf{n}_{j}^{\ast}   \vert,
\end{eqnarray}
to compare the analytic $\mathbf{a}_{j}$ and numeric $\mathbf{n}_{j}$ normal modes of the nine waveguide realization in Fig. \ref{fig:Fig3}(b). 
A fidelity value of one points to identical vectors, while a zero value to orthogonal vectors.
The mean average value for the fidelities in our example is $0.913 \pm 0.059$ points to good agreement that can be improved between our analytic and finite element models. 
We want to emphasize that the lowest fidelities arise from the two pairs of normal modes with shared effective propagation constant. 
This points to the fact that it may be possible to construct a linear superposition for each of these pairs that has a better overlap with the closed form analytic modes.

%%%%%%%%%%%%%%%%%%%%%%%%%%%%%%%%%%%%%%%%%%%%%%%%%%%%%%%%%%%%
\section{Forcing Zero-Energy modes} \label{sec:Sec4}

It is straightforward to realize that the limit case,
\begin{eqnarray}
\epsilon \rightarrow \beta_0-\beta_{\mathrm{min}},
\end{eqnarray}
forces a pseudo zero-energy mode in the mode coupling matrix partner.
Doing so invalidates the Cholesky decomposition SUSY results as the reduced mode coupling matrix arising from this choice is not positive definite. 
Still, the Cholesky lower $\mathbf{A}$ and upper $\mathbf{A}^{\dagger}$ triangular pair reconstructs the original coupled mode matrix $\mathbf{M}$ and provides a partner $\mathbf{P}$ that has two pseudo zero-energy modes. 
One of these modes is an isolated localized mode uncoupled to the array and the other is a normal mode of the array.
However, the algebraic properties that sustain the SUSY analogy are not fulfilled; for example, the intertwining relations are no longer valid.

\begin{figure}[htbp]
	\includegraphics{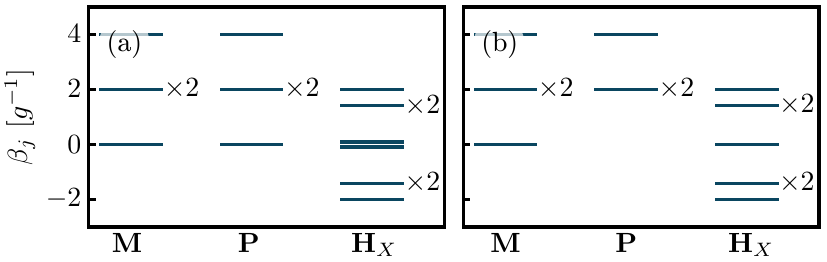}
	\caption{Propagation constants of the original coupled array $\mathbf{M}$, its partner $\mathbf{P}$ and its Cholesky square root array $\mathbf{H}_{X}$ for (a) broken SUSY and (b) forcing a pseudo zero-energy mode.}
	\label{fig:Fig4}
\end{figure}

As a practical example, let us discuss the Cholesky arrays for a four waveguide necklace. 
The simplest way to force a pseudo zero-energy mode is choosing the decomposition parameter $\epsilon = 2 g$  \cite{Smirnova2019p1120}.
This produces a null fourth column in the Cholesky matrix $\mathbf{A}$.
Physically, this means that the SUSY partner is a three-waveguide array that has identical normal-modes to the original mode coupling matrix but for the one corresponding to the lowest propagation constant; compare first two columns in Fig. \ref{fig:Fig4}(a) and \ref{fig:Fig4}(b).
In the square root lattice, this means that the eight waveguide becomes decoupled from the fourth waveguide, Fig. \ref{fig:Fig2}(b). 
Thus, instead of the original broken SUSY without a pseudo zero-energy mode, third column in Fig. \ref{fig:Fig4}(a), we do not account for the mode localized in the decoupled waveguide and obtain a spectrum with a null effective propagation parameter mode, third column in Fig. \ref{fig:Fig4}(b).
Formally, the arrays constructed in this manner do not fulfil SUSY QM.
For example, the pseudo zero-energy mode does not arise from SUSY considerations but for the fact that we have an effective odd-dimensional, real symmetric, traceless mode coupling matrix.

%%%%%%%%%%%%%%%%%%%%%%%%%%%%%%%%%%%%%%%%%%%%%%%%%%%%%%%%%%%%
\section{Conclusion} \label{sec:Sec5}

We showed that Cholesky factorization is a reliable method to construct broken SUSY isospectral and square root partners of photonic lattices described by coupled mode theory.
The mode coupling matrices designed in this form fulfil all characteristics from SUSY QM with Witten index two.

We constructed the isospectral and square root partner of waveguide necklaces that may be experimentally realized using femtosecond laser-writting techniques.
Broken SUSY square root partners are interesting because negative coupling strengths arise for necklaces of dimension four or more. 
We used an additional waveguide to simulate such processes.
Comparison of our analytic predictions with numeric finite element model  simulations show good agreement in both cases.

It is possible to force a spectrum with reduced range that points to exact SUSY using reduced mode coupled matrices with null main diagonal. 
Although these are not positive definite as required by Cholesky factorization, the resulting Cholesky matrices provide feasible partner photonic lattices.
These partners do not correspond to exact SUSY as the intertwining relations are not fulfilled. 

%%%%%%%%%%%%%%%%%%%%%%%%%%%%%%%%%%%%%%%%%%%%%%%%%%%%%%%%%%%%
\begin{acknowledgments}
B.M.R.-L. acknowledges fruitful discussions with B. Jaramillo \'Avila and F.H. Maldonado Villamizar.
P.I.M.B. thanks B. Jaramillo \'Avila support with figure formatting.
\end{acknowledgments}

%%%%%%%%%%%%%%%%%%%%%%% References %%%%%%%%%%%%%%%%%%%%%%%%%
%\bibliography{references}

\end{document}